\documentclass[aps,showpacs,preprintnumbers,amsmath,amssymb,nofootinbib,preprint]{revtex4}

\usepackage{graphicx}
\usepackage{color}
      
\def \be  {\begin{equation}}
\def \ee  {\end{equation}}
\def \ee  {\end{equation}}
\def \bea {\begin{eqnarray}}
\def \eea {\end{eqnarray}}

\begin{document}

\preprint{ECTP-2012-11\hspace*{0.5cm}and\hspace*{0.5cm}WLCAPP-2013-08}

\title{An Estimate of the Thermodynamic Pressure in High-Energy Collisions}

\author{Abdel Nasser~Tawfik\footnote{http://atawfik.net/}}
\affiliation{Egyptian Center for Theoretical Physics (ECTP), Modern University for Technology and Information (MTI), 11571 Cairo, Egypt}
\affiliation{World Laboratory for Cosmology And Particle Physics (WLCAPP), Cairo, Egypt}

\date{\today}


\begin{abstract}

We introduce a novel approach to estimate the thermodynamic pressure from heavy-ion collisions based on recently measured higher-order moments of particle multiplicities by the STAR experiment. We start with fitting the experimental results in the most-central collisions. Then, we integrate them back to lower ones. For example, we find that the first-order moment, the mean multiplicity, is exactly reproduced from the integral of variance, the second-order moment. Therefore, the zero-order moment, the thermodynamic pressure, can be estimated from the integral of the mean multiplicity. the possible comparison between such a kind of pressure (deduced from the integral of particle multiplicity) and the lattice pressure and the relating of Bjorken energy density to the lattice energy density are depending on lattice QCD at finite baryon chemical potential and first-principle estimation of the formation time of the quark-gluon plasma (QGP).  

\end{abstract}


\maketitle


\section{Introduction}
\label{sec:intr}

The structure of matter at pressure and/or energy density much larger than the critical values defining the hadron-quark deconfinement phase transition(s), is to a large extend well understood. One of the yet-unsettled problems of theoretical physics is the characterization of the equation(s) of state (EoS) describing the behaviour of thermodynamic quantities at finite temperatures and densities. To this end, we utilize effective models, lattice QCD simulations and high-energy collisions. To the latter, the ultimate experimental evidence is coupled. On one hand, the lattice QCD calculations turn to be very reliable, especially at high temperatures and densities. On the other hand, the exact EoS of the hadronic matter is still rather complicated to be modelled. In describing the ground state properties of nuclear matter having large number of finite nuclei, the Hartree-Fock theories using Skyrme effective interactions are found quite successful \cite{hf13a,hf13b,hf13c}. Nevertheless, it seems that serious concerns about basic physical symmetries arise, when using EoS derived from the Skyrme interactions \cite{shortcoming}, especially at finite temperatures. 

The early Universe is assumed to undergo a rapid phase transition from a phase dominated by colored degrees of freedom (partons) to a phase dominated by color-neutral degrees of freedom (hadrons) \cite{Tawfik:2011sh}. Such a phase transition is now routinely reproduced in heavy-ion collisions at the Relativistic Heavy Ion Collider (RHIC) and the Large Hadron Collider (LHC). The total number of produced particles exceeds several thousands. Therefore, one can expect that the generated system contains collective phenomenon. In light of this, we recall that the hydrodynamic description of the heavy-ion collisions goes back to Landau in early fifties of the last century. The hydrodynamics predicts long-range correlations induced by thermal noise.

The statistical-thermodynamic quantities, like pressure, energy density and number density, are well-known tools to describe nature, degrees of freedom, decomposition, size and even the overall dynamics controlling evolution of the medium from which they are originating \cite{Tawfik:2014eba}. The first-principle lattice calculations offer an excellent control on these quantities, especially at equilibrium. The effective models based on various approximations are also able to shed light on these quantities. The present work suggests a novel approach towards deducing the thermodynamic pressure from the high-energy collisions with different centralities. {\it We would like to highlight that this estimate is approximative}.  As stated in the title, the proposed method is {\it an estimate} due the various sources of uncertainties, which will be discusses in sec. \ref{sec:rmks}. The success in estimating the thermodynamic pressure opens horizon for other thermodynamic quantities, such as the energy density which can be compared with the Bjorken energy density. Based on the higher-order moments of particle multiplicity measured by STAR at RHIC, section \ref{sec:hm1}, we propose to estimate the pressure. 

It is in order now to highlight possible procedure in estimating energy density, for instance. This is not a straightforward implementation of the proposed method. Estimating the thermodynamic pressure which is directly related to the free energy, the energy density can be deduced from the derivative of the free energy with respect to inverse temperature. This obviously illustrates the need to implement another variable different than the one responsible for the derivatives of the high-order moments, the chemical potential.  In a future work will analyse the impact of this new variable. 

Reliable estimation of EoS (statistical-thermodynamic quantities) has deep impacts on various cosmological and astrophysical aspects. For instance, viscous EoS deduced from lattice QCD and heavy-ion collisions was implemented in early Universe (QCD era)  \cite{Tawfik:2011sh,Tawfik:2011mw,Tawfik:2010pm,Tawfik:2010ht,Tawfik:2010bm,Tawfik:2009mk,Tawfik:2010mb}.  Section \ref{sec:do} is devoted to the discussion and outlook. Our remarks about the proposed methis are elaborated in section \ref{sec:rmks}.

\section{Higher-order moments of net-proton multiplicity}
\label{sec:hm1}

The higher-order moments can be studied in different physical quantities. For example, the higher-order moments of charged-particle multiplicity distribution have been predicted four decades ago \cite{gupta72}. Recently, the higher-order moments have been reported by STAR at various RHIC energies \cite{star1a,star1b,star1c,star2a,star2b} and by lattice QCD calculations \cite{lqcd1,lqcd1a,lqcd2}. The empirical relevance of the higher-order moments to the experimental measurements has been suggested in Ref. \cite{endp1}. Accordingly, the measurement of the correlation length $\xi$ is very crucial. It is bounded by the finite volume and  lifetime of the system through causality principle. In a relativistic system, the growth rate of correlated domains should be ultimately bounded by $c$.

The normalization of higher-order moments gives additional insights about their properties. From statistical point of view, the normalization is done with respect to the standard deviation $\sigma$, which is related to $\xi$, i.e. $r-$th order moment is to be normalized to $\sigma^{r}$. Therefore, it provides with a tool relating moments with various orders to the experimental measurement. For instance, the susceptibility of the distribution, second order moment, gives a direct measure for $\sigma$. 

There are several techniques to scale the correlation functions. The survey system's optional statistics module represents the most common technique, i.e. Pearson or product moment correlation. This module includes the so-called partial correlation which seems to be useful when the relationship between two variables should be highlighted, while effect of one or two other variables can be removed. The products of higher-order moments of conserved quantities can be directly connected to the corresponding susceptibilities and be directly related to long range correlations \cite{endp5,qcdlike,latqcd1}.

Some details about the experimental procedure are now in order. The STAR detector excellently identifies produced particles and has a large acceptance for the event-by-event fluctuation analysis. For each beam energy, an extensive quality assurance is performed  in order to minimize the fluctuation of detector efficiency. From the Time Projection Chamber (TPC) center, the event-by-event net-proton distribution lying within $\pm 30~$cm along the $z$ position of the interaction point  and $2.0~$cm radius in the transverse plane can be excellently detected.

The geometry of each heavy-ion collision (and thus centrality of the collision) can be characterized by the number of participating nucleons $\langle N_{part}\rangle$. By using centrality bin width correction \cite{cw1,cw2}, finite centrality bin width effect has been avoided. The charged produced particles can measured between $0.2<p_T<2.0~$GeV and  $|\eta|<0.5$, where $p_T$ and $\eta$ are transverse momentum and pseudo-rapidity, respectively. Centrality selection has been done by uncorrected charged particles measured within the wider window $0.5<|\eta|<1.0$ in order to avoid the auto-correlation effect in the higher moments analysis.

The positive and negative charged particles distributions are assumed to be independently Poisson distribution where no dynamical correlation among the positive and negative charge particles are taken into account. Accordingly, the net-proton distribution is taken as a Skellam distribution \cite{skellm}, which is a discrete probability distribution of the difference between two statistically independent  variables. Each of these {\it random} variable has Poisson distribution and a different expected value. The Skellam distribution  is baseline for this analysis \cite{Sahoo}.

\subsection{Integrals of higher-order moments}
   
\begin{figure}[htb]
\centering{
\includegraphics[angle=-90,width=7.cm]{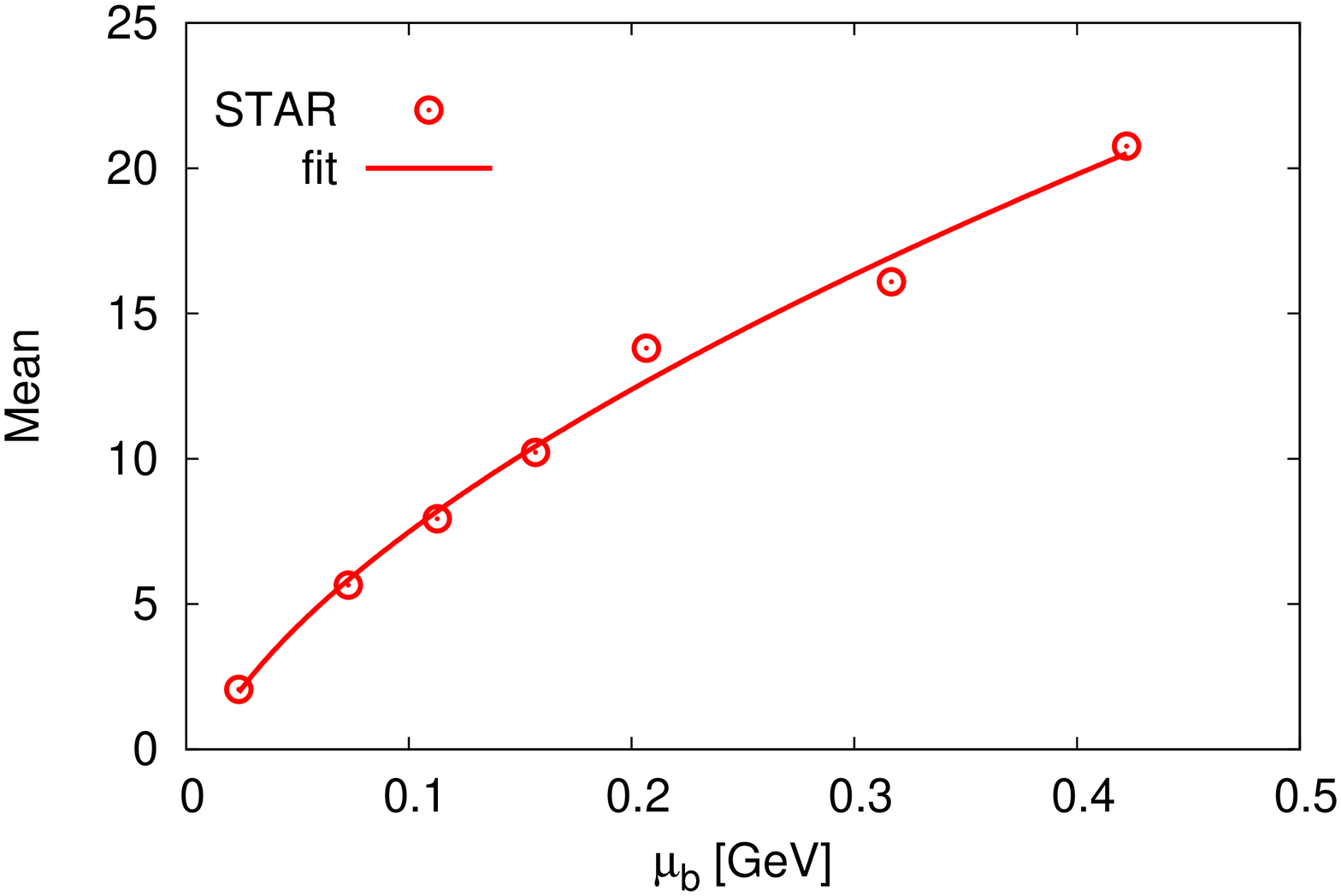}
\includegraphics[angle=-90,width=7.cm]{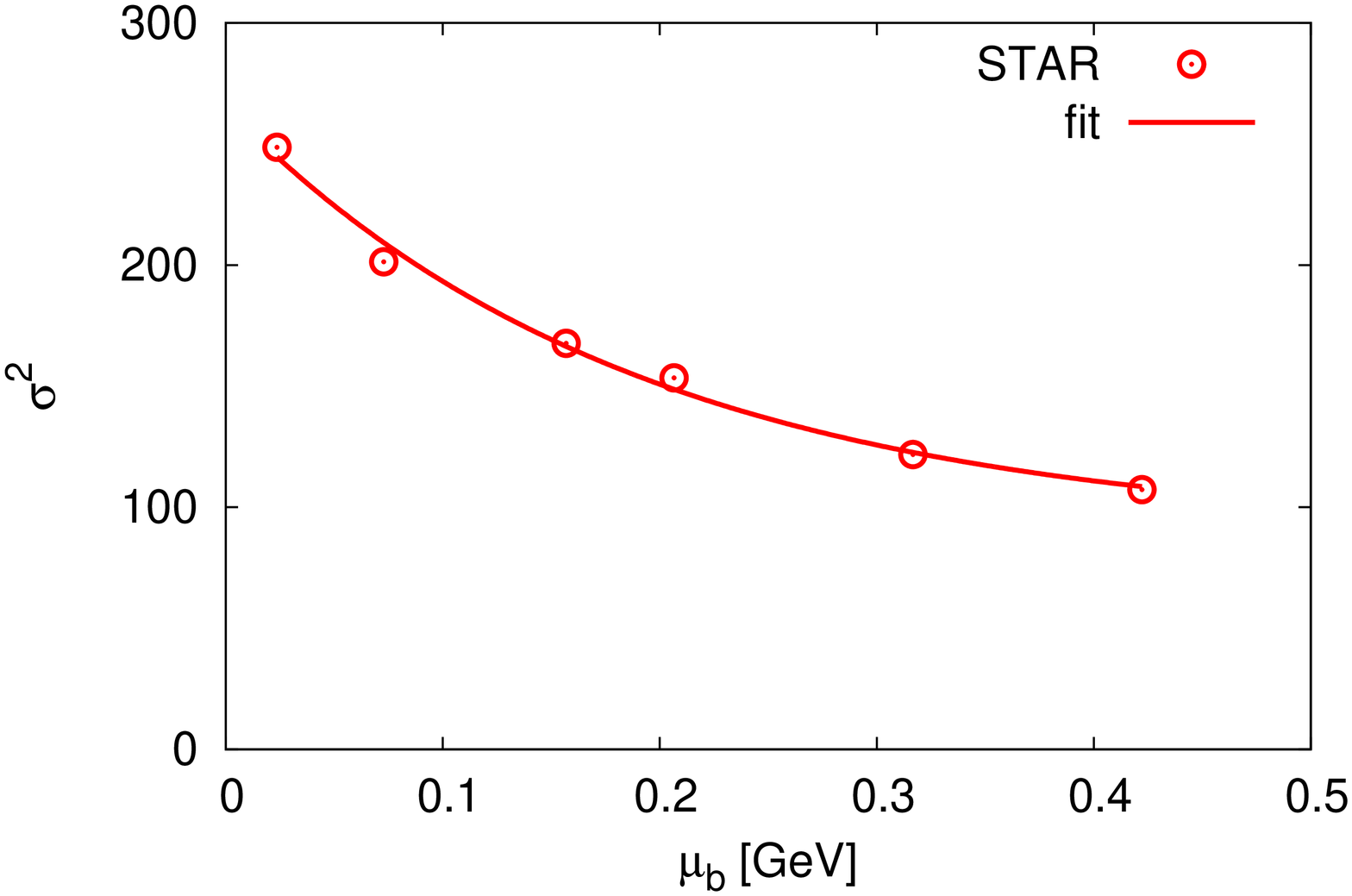} 
\caption{Mean (left panel) and standard deviation (right panel) are plotted as function of the baryon chemical potential $\mu_b$. The experimental results (symbols) are taken Ref. \cite{Sahoo}. The curves represent Eqs. (\ref{eq:meann}) and (\ref{eq:sigmm}), respectively. }
\label{fig:myMS} 
}
\end{figure}

\begin{figure}[htb]
\centering{
\includegraphics[angle=-90,width=7.cm]{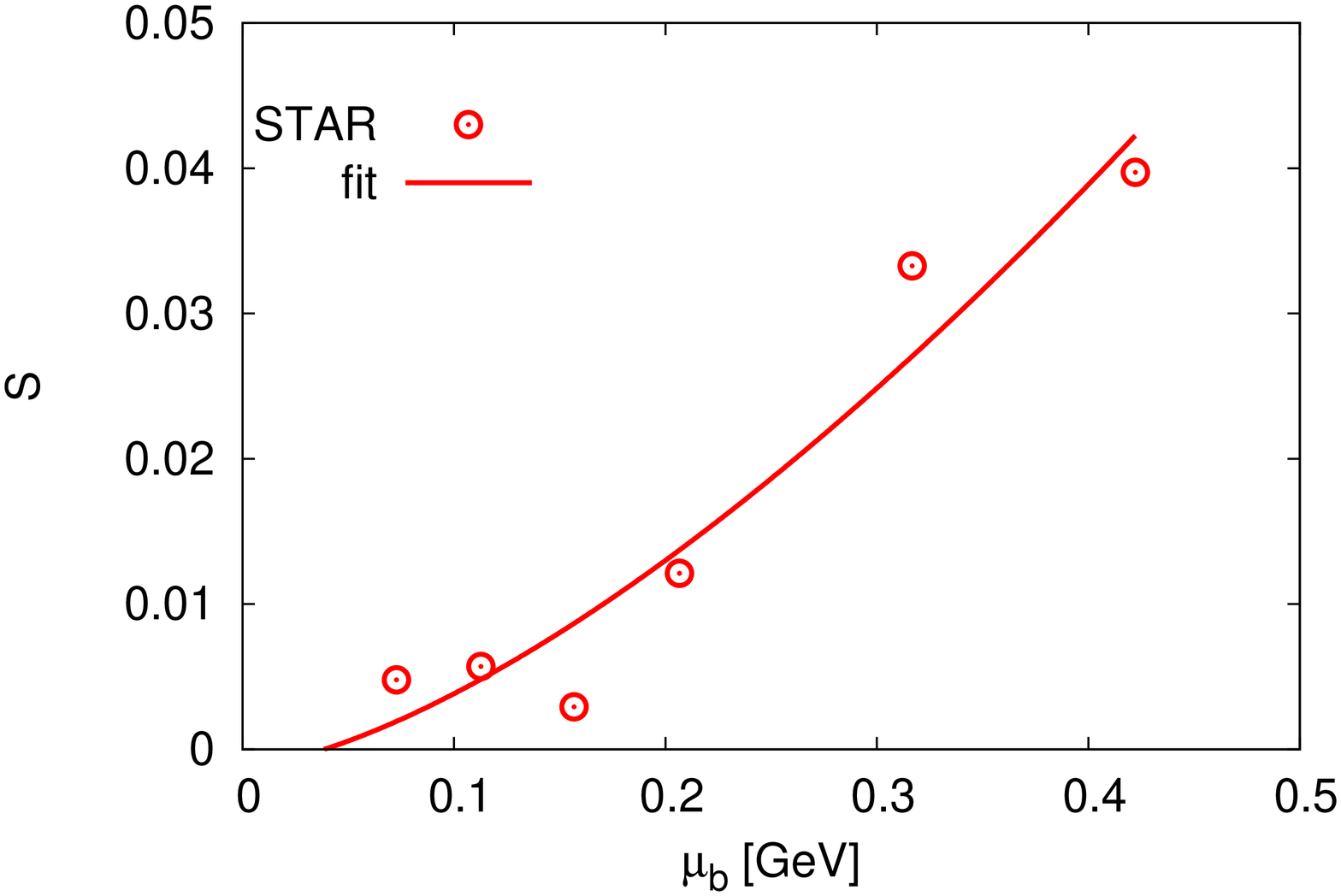}
\includegraphics[angle=-90,width=7.cm]{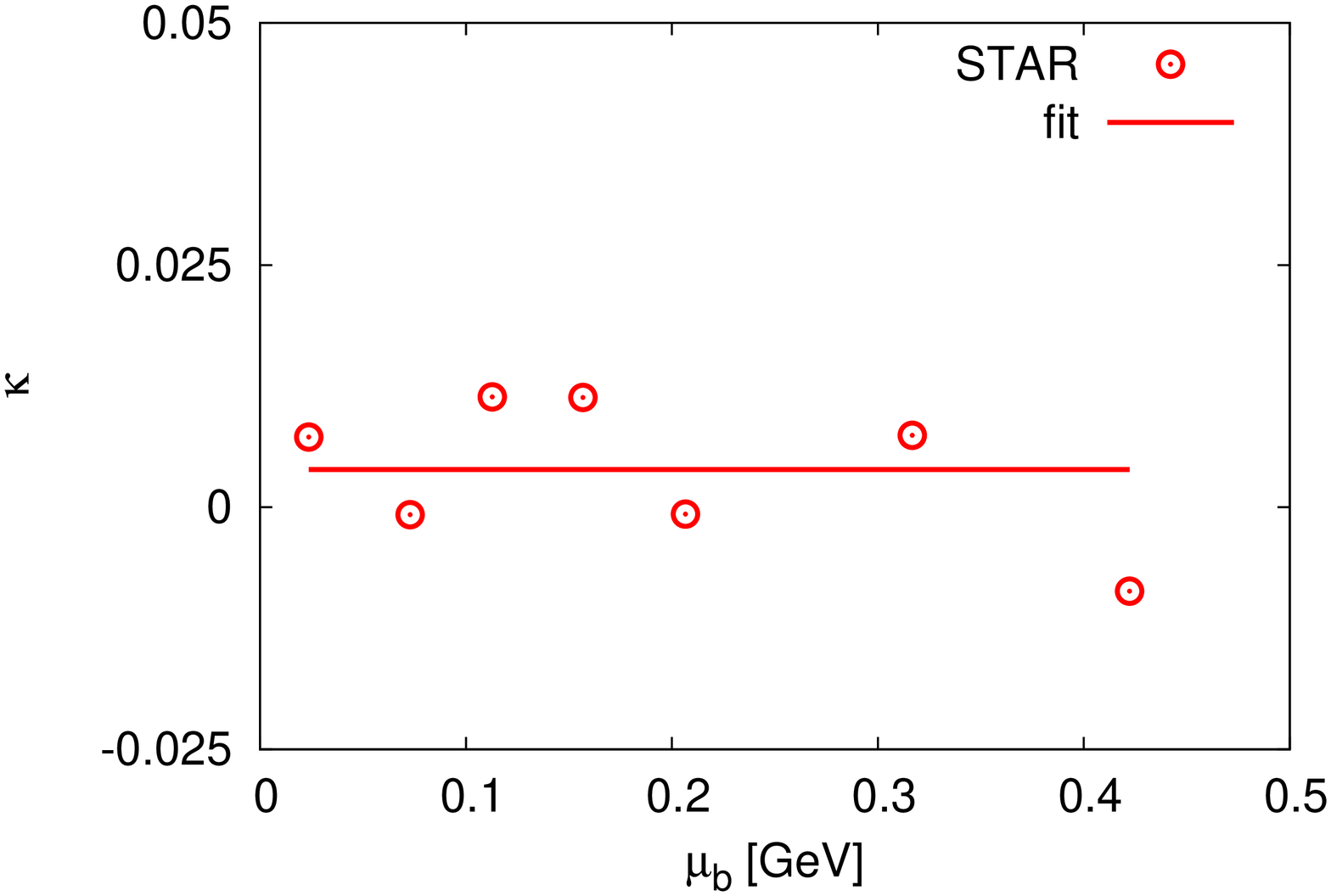}
\caption{Skewness (left panel) and kurtosis (right panel) are plotted with respect to the baryon chemical potential $\mu_b$. The experimental results (symbols) are taken from Ref. \cite{Sahoo}. The curves represent Eqs. (\ref{eq:skewnesss}) and (\ref{eq:kurtosiss}), respectively. }
\label{fig:mySK} 
}
\end{figure}

The experimental higher-order moments at the largest $
\langle N_{part}\rangle$ are related to most-central high-energy collisions and likely comparable to lattice QCD simulations. The effects of  varying centralities of the collisions shall be discussed in section \ref{sec:cd1}. The first four moments of net-proton multiplicity are plotted with the baryon chemical potential $\mu_b$, which is related to the nucleon-nucleon center-of-mass energy $\sqrt{s_{NN}}$ \cite{jean2006}. For simplicity, we write $\sqrt{s}$ instead of $\sqrt{s_{NN}}$.
\bea
\mu_b &=& \frac{1.31\pm 0.03\, [\text{GeV}]}{1+(0.27\pm 0.01\, [\text{GeV}^{-1}]) \sqrt{s_{NN}}}. \label{eq:sqrtsmus}
\eea
We notice that $M$ and $S$ increase with increasing $\mu_b$, while the standard deviation $\sigma$ decreases and kurtosis $\kappa$ remains almost constant. The given curves represent the statistical fits, Eqs. (\ref{eq:meann})-(\ref{eq:kurtosiss}). The baryon chemical dependence of mean, standard deviation, skewness and kurtosis, respectively, reads  
\bea
M &=&  a_1 \mu_b^{b_1} + c_1, \label{eq:meann}\\
\sigma^2 &=& a_2 \mu_b^{b_2} + c_2, \label{eq:sigmm} \\
S &=&  a_3 \mu_b^{b_3} + c_3, \label{eq:skewnesss}\\
\kappa &=& 0.004\pm 0.002, \label{eq:kurtosiss}
\eea
where $a_1=37.8\pm 0.11$, $b_1=0.6\pm 0.11$, $c_1=-2.105\pm 1.1$, $a_2=175.75\pm 11.07$, $b_2=-5.27\pm 1.02$, $c_1=89.5\pm 13.13$. $a_3=0.16\pm 0.06$, $b_3=1.5\pm 0.5$ and $c_3=-0.001\pm 0.0005$.  Equations (\ref{eq:meann})-(\ref{eq:kurtosiss}) are analytical expressions representing the first four orders of moments. The procedure introduced in the present work is the integration of the moments of particle multiplicity. This consecutively results in moments with lower orders. By implementing this we propose that the thermodynamic pressure can be estimated.

\subsection{Thermodynamic pressure in high-energy collisions}

Our approach towards deducing thermodynamic pressure from the higher-order moments measured in the heavy-ion collisions is based on:
\begin{itemize}
\item the higher-order moments of particle multiplicities are accessible in the high-energy experiments through the correlation length $\xi$,
\item fitting the experimental results on higher-order moments makes it feasible to integrate them back to lower ones, and
\item therefore, the zero-order moment of {\it mean} particle multiplicities, the intensive thermodynamic pressure, can be obtained, accordingly.   
\end{itemize}

\begin{figure}[htb]
\centering{
\includegraphics[angle=-90,width=8.cm]{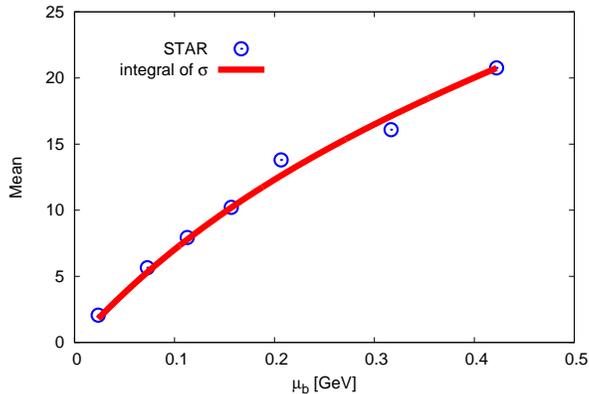}
\caption{The STAR results on normalized mean multiplicity (STAR BES I) are given in dependence on the baryon chemical potential, $\mu_b$. The curve represents the integral of Eq. (\ref{eq:sigmm}), left-hand panel of Fig. \ref{fig:myMS}.  }
\label{fig:intgrdMean} 
}
\end{figure}

\begin{figure}[htb]
\centering{
\includegraphics[angle=-90,width=7.cm]{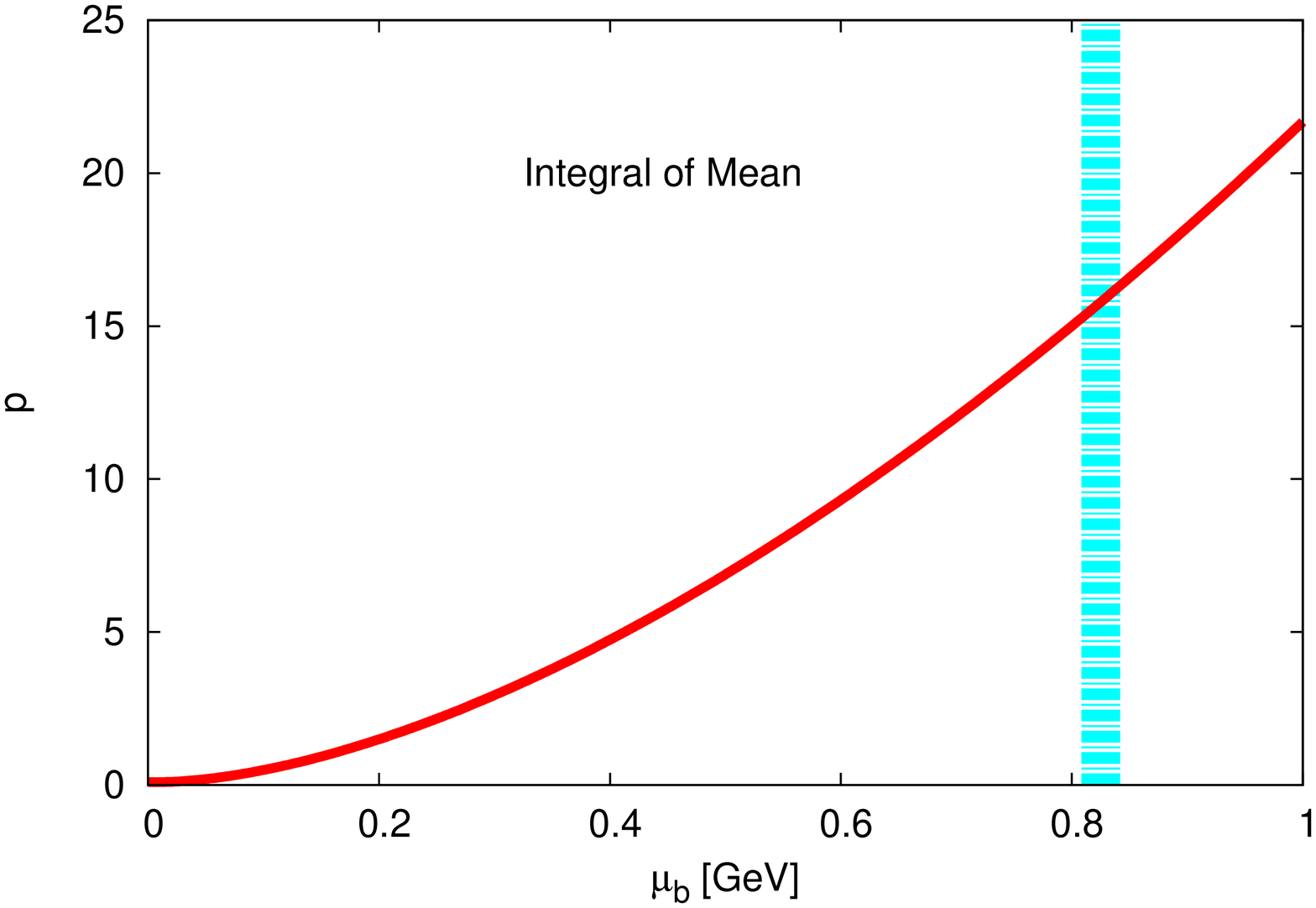}
\includegraphics[angle=-90,width=7.cm]{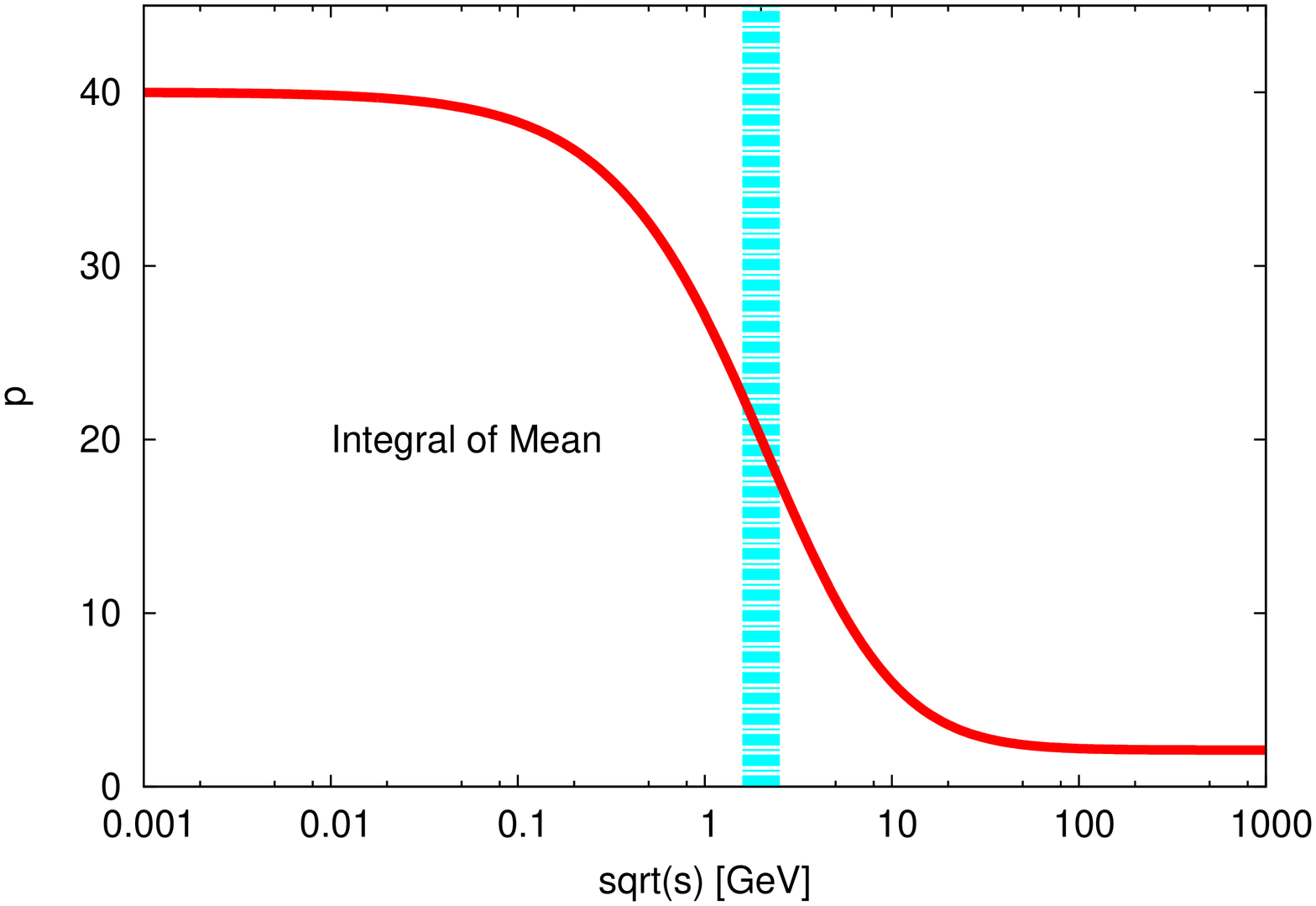}
\caption{The integrals of mean multiplicity distribution is given in dependence on the baryon chemical potential (left) and center-of-mass energy (right). The vertical bands show the critical deconfinement phase transition. The overall behavior has is nearly 'tangen-hyperbolicus'-like.}
\label{fig:intgrdPrssr} 
}
\end{figure}

In order to test the given approach, we can apply it on the variance distribution, for example. The integral of the variance gives an estimation for the mean multiplicity. In Fig. \ref{fig:intgrdMean}, the STAR results on normalized mean multiplicity (symbols) are given as function of the baryon chemical potential, $\mu_b$. The translation between $\sqrt{s}$ and $\mu_b$ is governed by Eq. (\ref{eq:sqrtsmus}). The integral of variance, Eq. (\ref{eq:sigmm}) and left-hand panel of Fig. \ref{fig:myMS}, is represented by the thick solid curve, 
\bea
M(\mu_b) &=& \alpha_1 \mu_b + \frac{\beta_1}{\gamma_1} \exp(\gamma_1\, \mu_b) + \delta_1, \label{eg:intvrnc1}
\eea
where $\alpha_1=25.94\pm3.8$, $\beta_1=58.6\pm 3.7$, $\gamma_1=-5.27\pm1.024$ and $\delta_1=11$. It is quite obvious that the two curves in Fig. \ref{fig:intgrdMean} and left-hand panel of Fig. \ref{fig:myMS} are likely identical. This makes it eligible to integrate the mean multiplicity, Eq. (\ref{eq:meann}) right-hand panel of Fig. \ref{fig:myMS}, in order to deduce the thermodynamic pressure, Fig. \ref{fig:intgrdPrssr}.  As expected, the overall behavior is monotonic. Increasing $\mu_b$ leads to an increase in the resulted pressure. This behavior was observed in lattice QCD at finite $\mu_b$ \cite{fodor2012}. When $\mu_b$ is translated into $\sqrt{s}$, the dependence of integral of mean or double integrals of variance (dimensionless pressure), Eq. (\ref{eg:intvrnc1}), on $\sqrt{s}$ is given in right panel. Here, the behavior is non-monotonic. It is worthwhile to notice that the relation $\mu_b(\sqrt{s})$ was phenomenologically estimated in the final state of particle production.

At small $\sqrt{s}$, the resulting pressure is large. This is obviously related to the large stopping power related to the incident energy. When the value of $\sqrt{s}$ reaches about $1-2~$GeV, the pressure considerably decreases. This would be related to a drastic change in the effective degrees of freedom. At large $\sqrt{s}$, the resulting pressure gets relatively small values. Further raise in  $\sqrt{s}$ does not change the thermodynamic pressure. The overall behavior can be described as 'z-shape' or nearly 'tangen-hyperbolicus'-like.

The integral of mean multiplicity (dimensionless thermodynamic pressure) reads
\bea
p(\mu_b) &=& a_5 \mu_b+\frac{b_5}{c_5+1} \mu_b^{c_5+1} + d_5, \\
p(\sqrt{s}) &=& \frac{a_6}{2} s + \frac{b_6}{c_6^2} \exp\left(c_6 \sqrt{s}\right) +d_6 \sqrt{s},
\eea
where $a_5=-2.1\pm1.2$, $b_5=37.8\pm1.97$, $c_5=0.6\pm0.011$, $d_5=11$, $a_6=89.5\pm1.3$, $b_6=58.6\pm3.9$, $c_6=-5.27\pm0.02$, and $d_6=11$. This result is obviously original. So-far this quantity was not accessible in the high-energy experiments. An extension to other thermodynamic quantities is therefore planned in future works.

To make the new method more clear we summarize the procedure, shortly:
\begin{itemize}
\item From consecutive integrals of the high-order moments, an analytical expression for the number density (the multiplicity) can be deduced.
\item By integrating this expression with respect to the chemical potential, the thermodynamic pressure will be estimated. 
\end{itemize}

\subsection{Remarks on proposed estimate}
\label{sec:rmks}

We want to highlight that the present work does not claim that this is an exact method to deduce the thermodynamic pressure. As we discussed earlier, to this end, we still have to take into consideration various uncertainties. \begin{itemize}
\item First, the direct comparison of the pressure (deduced from integral of mean multiplicity, especially in central collisions) with the lattice pressure is not possible. 
\item Second, the comparison of the Bjorken energy density with the lattice energy density is also not allowed. An estimate of the thermodynamic energy density from the high-energy collisions seems to implement sophisticated methods.
\item Third, we should verify the assumption that the moments of net-protons are almost equivalent to the ones of all baryons should be correct.   Nevertheless, the assumption that the net-proton higher-order moments are nearly equivalent to that of net-baryon finds supporters \cite{nxu}.
\item Fourth, seeking for completeness, we recall the electric charge distribution. It is conjectured that their moments are related to the Taylor expansion coefficients of the pressure with respect to electric charge chemical potential at the freeze-out. This would be insufficient to determine the pressure, since the overall integration constant cannot be fixed. 
\item Fifth, the experimental measure for the higher moments utilizes particle multiplicity and very rarely the correlation length. Thus, the possible highlights on the  critical phenomena related to hadron-quark phase transition(s) are apparently limited to statistical insights.
\end{itemize}

 In relating our estimate of the thermodynamic pressure to that deduced from lattice QCD simulations, the difficulty of performing lattice calculations at finite baryon chemical potential $\mu_b$ should be taken into account. Assuming chemical freeze-out \cite{Tawfik:2013eua,Tawfik:2013dba,Tawfik:2013pd,Tawfik:2005qn}, the latter is to be related to the center-of-mass energy $\sqrt{s}$. 

In relating our estimate of the thermodynamic pressure to the hydrodynamics, the formation time $\tau$ should be measured.  The guess that $\tau$ varies with  $\sqrt{s}$ should be first verified, before confronting and/or merging pressure and energy density measured in high-energy collisions and calculated in lattice QCD. So far, there is no direct measurement or first-principle estimation for the QGP formation time $\tau$. In a very simple treatment,  the formation time of partons is indirectly related to their average energy \cite{formt}.  The hadrons are emitted only from the surface of freeze-out \cite{Tawfik:2013eua,Tawfik:2013dba,Tawfik:2013pd,Tawfik:2005qn}, whereas direct photons are conjectured to be emitted throughout the life-time of the evolving system. Therefore, electromagnetically and/or weakly interacting probes would carry information about the formation time. For the QGP temperature, we have so-far two different measurements using direct photons.  Further measurements are very desired. Also the estimation of the unified energy density and its relation to the critical temperature using traverse mass of different particles should be extended.

\subsection{An estimate of the higher moments}
\label{sec:cd1}

\begin{figure}[htb]
\centering{
\includegraphics[angle=-90,width=7.cm]{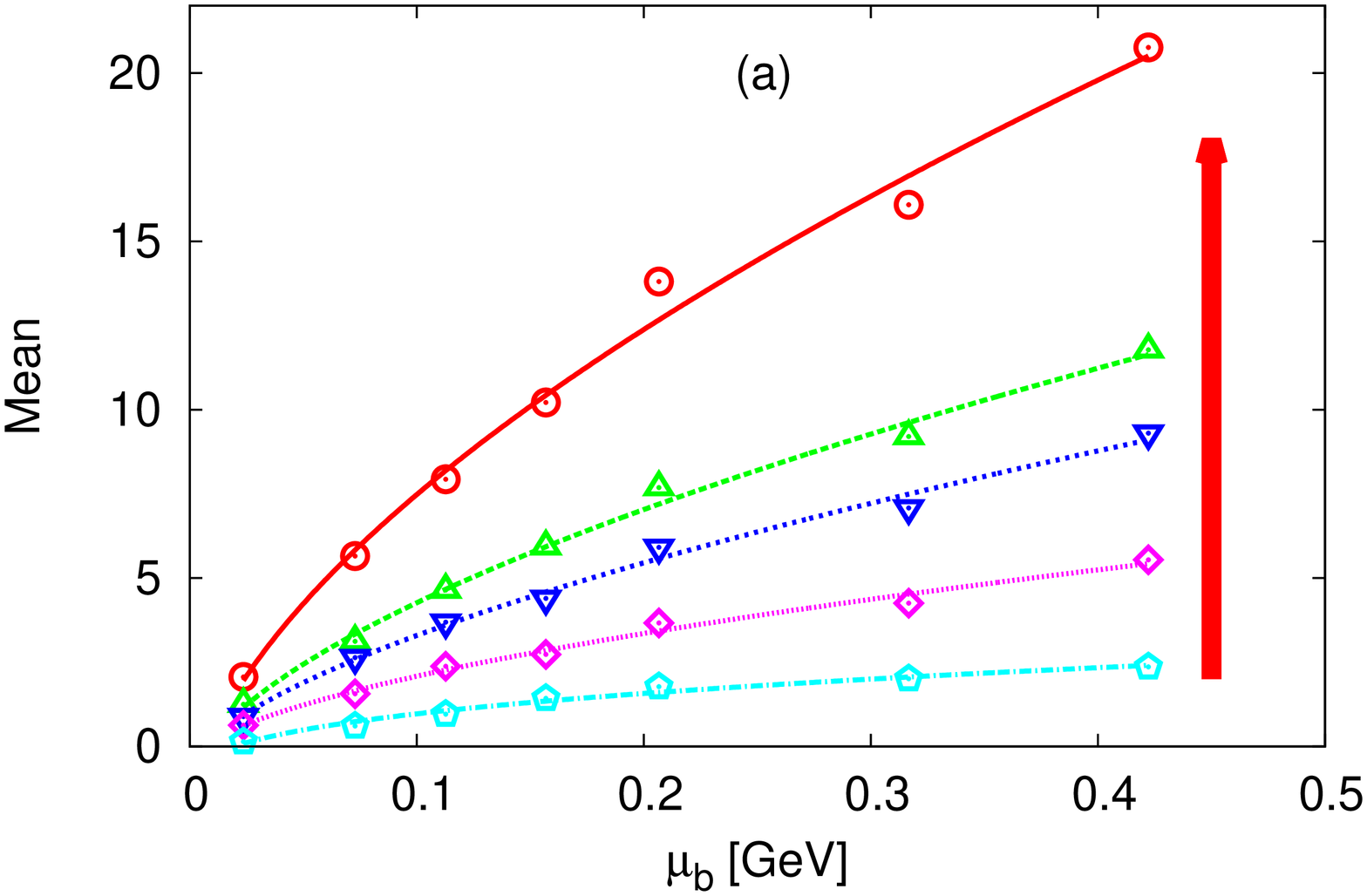}
\includegraphics[angle=-90,width=7.cm]{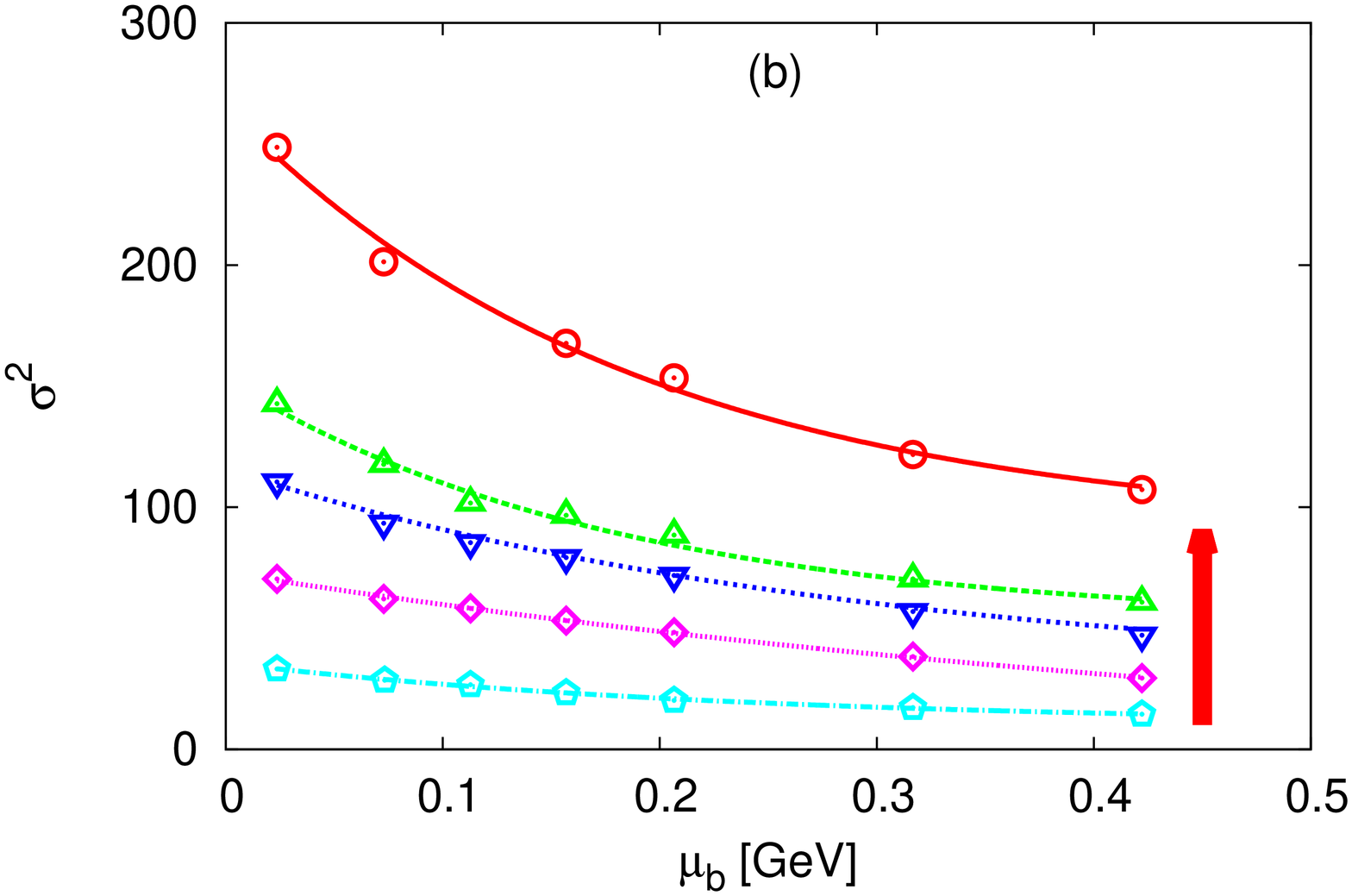} \\
\includegraphics[angle=-90,width=7.cm]{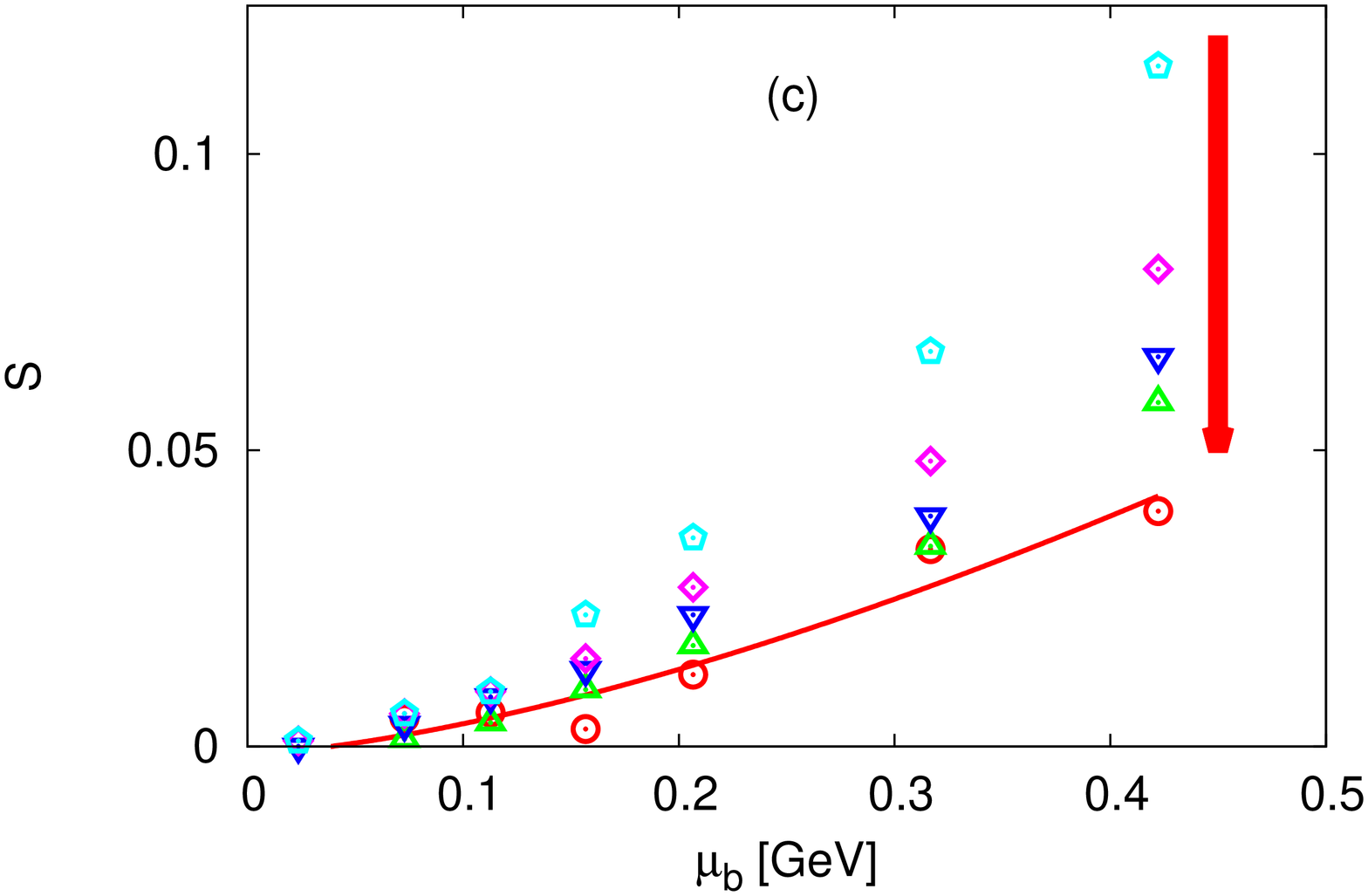}
\includegraphics[angle=-90,width=7.cm]{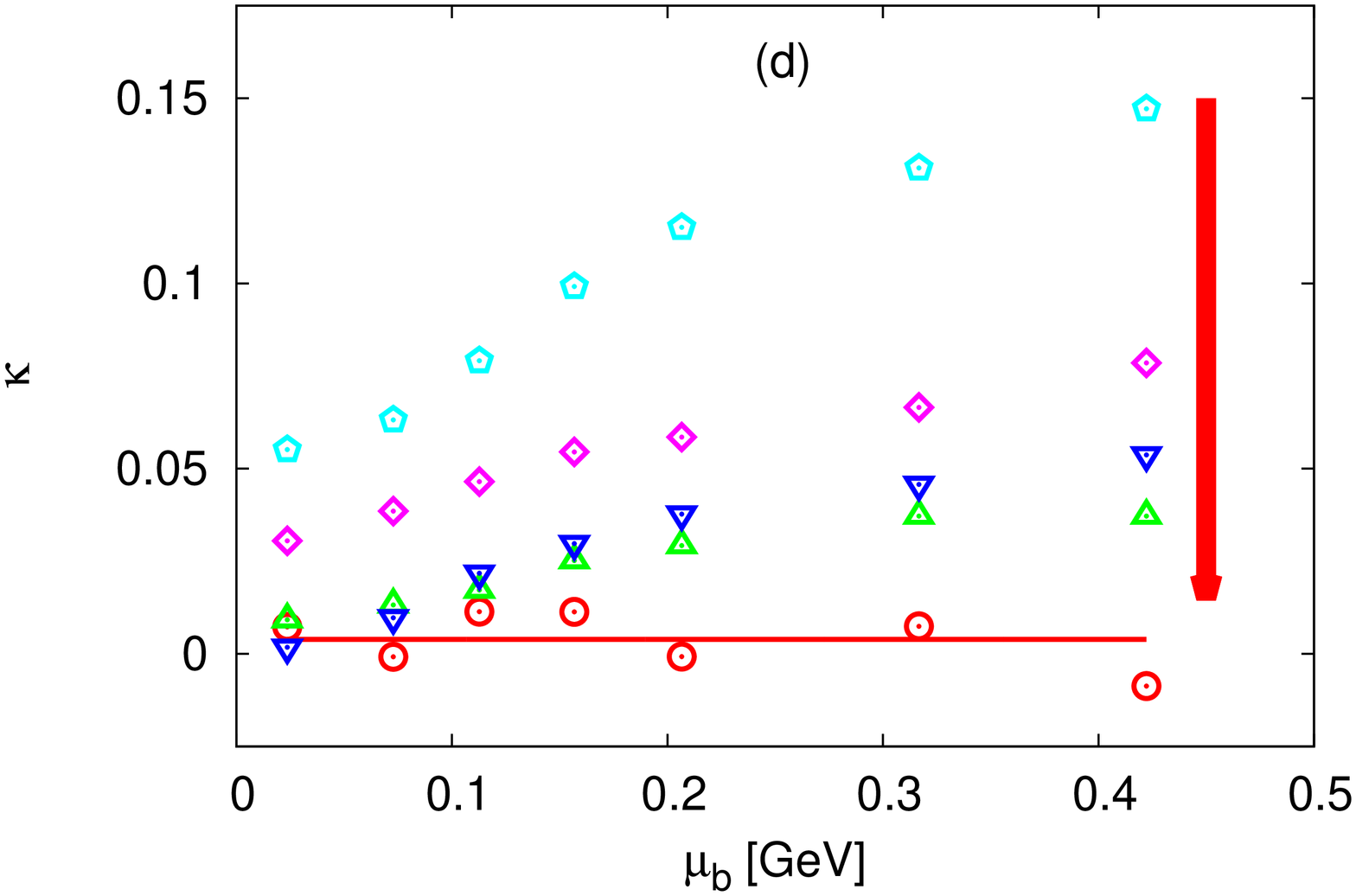} 
\caption{Mean (a), variance (b), skewness (c) and kurtosis (d) are given dependence on the collision centrality. Increasing centrality (marked by vertical arrow) is accompanied by increasing mean and variance and by decreasing skewness and kurtosis. Solid curves present the fits of the most-central results as given in Figs \ref{fig:myMS} and \ref{fig:mySK}. For mean, the fits for different centralities are shown. }
\label{fig:HMcntrl} 
}
\end{figure}

Fig. \ref{fig:HMcntrl} shows the dependence of the first four moments on centrality of the collision, which is geometrically defined by the averaged number of interacting nucleons $\langle N_{part}\rangle$. As given in previous sections, the maximum $\langle N_{part}\rangle$ is likely related to the most-central collisions. Varying $\langle N_{part}\rangle$ from $50$ to $200$, we move from peripheral to central collisions. We notice that increasing centrality (here marked by vertical arrow) is accompanied by increasing mean and variance and by decreasing skewness and kurtosis. The solid curves represent the fits of the most-central results as given in Figs \ref{fig:myMS} and \ref{fig:mySK}. Only for mean and variance, the fits for different centralities are shown. From these two quantities, we want to check again the validity of the given approach and then deduce the variation of the thermodynamic pressure with the collision centrality.

\begin{figure}[htb]
\centering{
\includegraphics[angle=-90,width=7.cm]{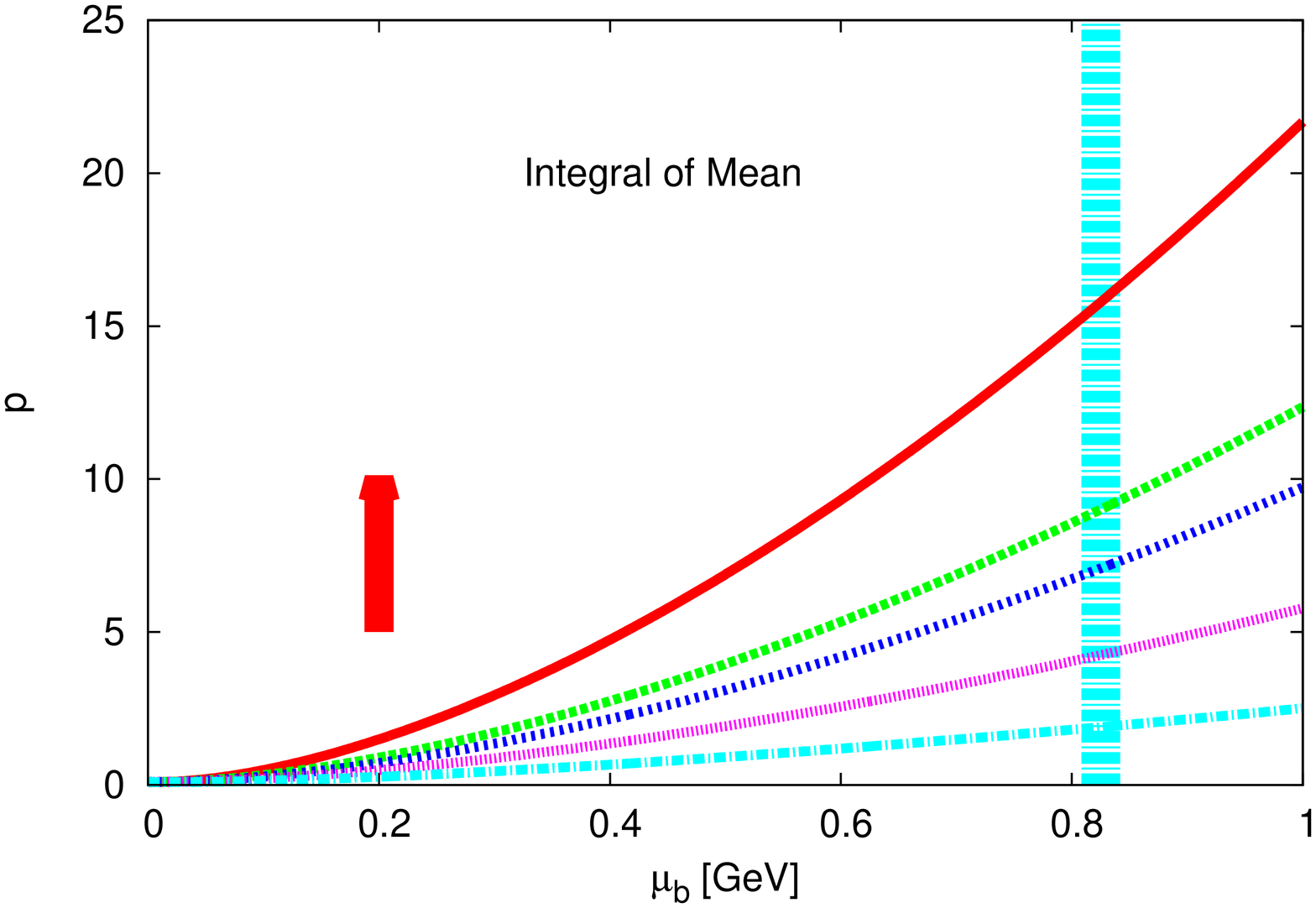}
\includegraphics[angle=-90,width=7.cm]{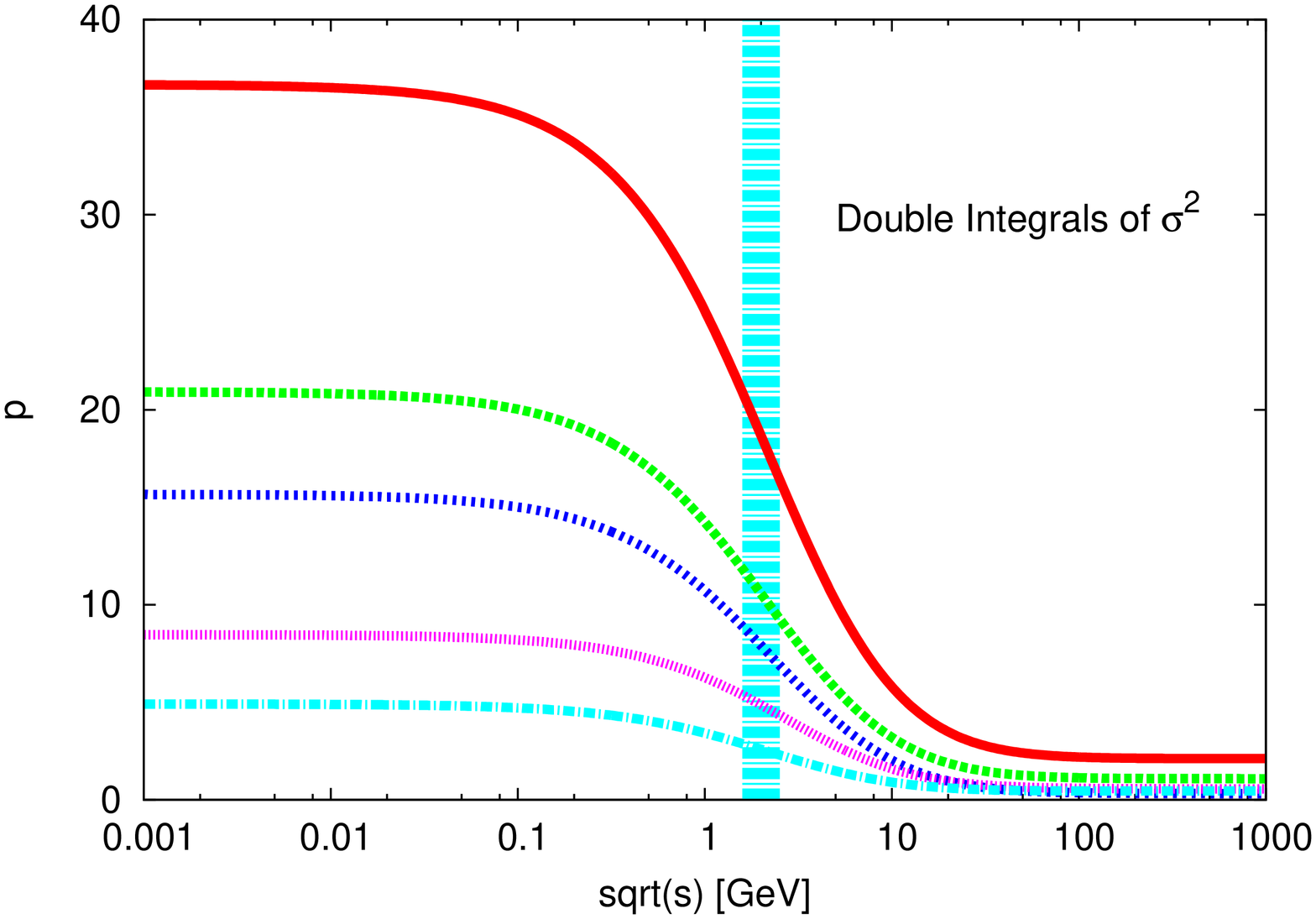}
\caption{Left-hand panel: integral of mean multiplicity at various centralities is given in dependence on the baryon chemical potential. Right-hand panel: double integrals of variance at different centralities are given in dependence on center-of-mass energy. Assuming that the critical deconfinement phase transition does not depend on the centrality, one vertical band referring to that is drawn. }
\label{fig:intgrdPrssrCentrality} 
}
\end{figure}

Left-hand panel of Fig. \ref{fig:intgrdPrssrCentrality} shows integral of mean at various centralities in dependence on the baryon chemical potential. As expected, the overall behavior is monotonic. Reducing the centrality reduces the values of integrals of mean (pressure). Right-hand panel shows  the dependence of double integrals of variance at various centralities in dependence on center-of-mass energy. Here, the behavior is non-monotonic. The overall behavior can be described as follows. At small $\sqrt{s}$, the resulting pressure is large. The large stopping power is likely the reason. When $\sqrt{s}$ reaches $1-2~$GeV, the resulting pressure considerably decreases. Assuming that the critical deconfinement phase transition does not depend on the centrality, one vertical band referring to that is drawn. A drastic change in the effective degrees of freedom is likely. At large $\sqrt{s}$, the resulting pressure gets relatively small values. Further raise in  $\sqrt{s}$ does not change the pressure.

\section{Discussion}
\label{sec:do}

\begin{figure}[htb]
\centering{
\includegraphics[angle=-90,width=14.cm]{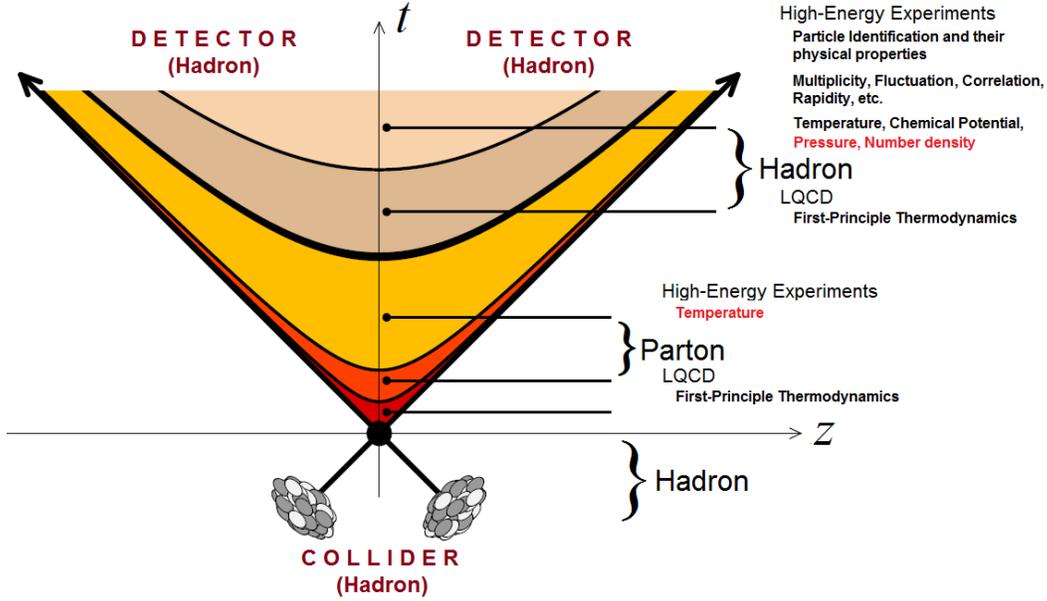}
\caption{Schematic illustration of various stages of a central heavy-ion collision. Measured and calculated thermodynamic quantities in partonic and hadronic phases are outlined.}
\label{fig:myStages} 
}
\end{figure}

Current particle detection technologies are mainly based on properties of the leptons and hadrons, for instance. In high-energy experiments, the inputs are hadrons (or leptons) accelerated to high energies and the outputs are produced colorless particles which finally freeze out, chemically and thermally, and then will be detected. The running strong coupling prevents quarks to escape from confined hadrons. These are the experimental obstacles against direct or even indirect detection of thermodynamic quantities in early stages of high-energy collisions, Fig \ref{fig:myStages}. 

The status of estimating the physical properties of hadronic and partonic matter is very shortly summarized in Fig. \ref{fig:myStages}. The first-principle lattice QCD calculations shed light on thermodynamics of the strongly interacting matter. Not only on numerical techniques the lattice calculations are depending, but also on the assumption of equilibrium statistical thermodynamics. The latter refers to a system that evolves to a state which is independent of the past history and time. The equilibrium state is specified by energy, volume and number of particles. For a system at equilibrium, there exists a positive differential entropy function which increases with energy at fixed volume and particle number. In other words, the entropy is only defined in an equilibrium system.  Non-equilibrium phenomena, such as heat conduction or diffusion, are apparently of great interest in high-energy physics. The non-equilibrium statistical mechanics distinguishes fast variables (characterized by microscopic time and space scales) from slow ones (characterized by macroscopic time and space scales) \cite{book}.

The physics program of high-energy experiments is very successful in deducing essential physical quantities like identification of produced hadrons, measuring their physical properties, counting their multiplicities, and estimating their fluctuations and correlations in different rapidity intervals, etc. The hadron multiplicities, fluctuations and correlations are experimental tools to analyse nature, composition, and size of medium from which they are originating. Based on analysing the particle abundances or momentum spectra, the degree of equilibrium of the produced particles can be estimated. The particle abundances can help to establish the chemical composition of the system. The momentum spectra can give additional information on the dynamical evolution and the collective flow. For instance, at chemical equilibrium, the entropy and number of produced particles are fixed. Thus, statistical-thermal approaches can be applied in order to estimate various thermodynamic quantities \cite{Tawfik:2006yq,Tawfik:2006dk,Tawfik:2012zp,Tawfik:2012si,Tawfik:2008ii}. The agreement with the lattice calculations was found excellent \cite{Tawfik:2012ty,Karsch:2003vd,Karsch:2003zq,Redlich:2004gp,Tawfik:2004sw}. Both are likely related to central collisions. Therefore, the QCD phase diagram \cite{Tawfik:2004sw} and freeze-out curves can be mapped out \cite{Tawfik:2013eua,Tawfik:2013dba,Tawfik:2013pd,Tawfik:2005qn}.

\section{Summary and outlook}
\label{sec:sr}

Based on the higher-order moments of particle multiplicities, which are experimentally accessible through the correlation length $\xi$. We started with fitting the experimental results, we introduced an approach to estimate the thermodynamic pressure from the heavy-ion collisions. Afterwards, we integrated them with respect to the chemical potential. This resulted in the lower-order  moments. For example, the first-order moment, which is the mean multiplicity, was exactly reproduced from the integral of the variance, Fig. \ref{fig:intgrdMean}, which is the second-order moment. Accordingly, the zero-order moment, which is noting but the thermodynamic pressure, was obtained from the integral of mean multiplicity, Fig. \ref{fig:intgrdPrssr}. The concerns about the certainties of the proposed method have been elaborated.

One of the ultimate goals of the comprehensive program of high-energy experiments is a reliable estimation for all equation of state and that of motion. So far, we measure collective properties of produced particles, such as multiplicities, correlations and fluctuations. Furthermore, the trustful order-parameters characterizing the phase transition are related to thermodynamic quantities. The lattice QCD simulations at finite temperature and density make use of such order parameter in determining the critical phenomena accompanying the deconfinement to the colored phase, the quark-gluon plasma. In the present, we introduce an estimate of the thermodynamic pressure from the measured multiplicities. Should the proposed results be complete and accurate, we shall be able to compare the experiments directly with the lattice QCD simulations. Obviously, this brings the high-energy experiments to a very bright horizon and enables tackling various still-unsolved problems, for example, orders of the deconfinement phase-transition and critical temperature and exponents. The latter are related to the universality classes of the system of interest.  For an accurate estimation of the thermodynamic quantities:
\begin{itemize}
\item quite gauge of the phase space,
\item comprehensive measure of the baryon multiplicity,
\item full identification of the produced particles,
\item exact determination of the fireball volume,
\item complete estimation of possible correlations and fluctuations,
\item characterizing the geometrical (spacial) and temporal evolution of the fireball,
\item complemented metering of the freeze-out boundaries
\item etc.
\end{itemize}
are crucially significant. No doubt that advanced experimental facilities are indispensably essential to fulfil these demands.

\section*{Acknowledgement}
The author is very grateful to the anonymous referee for her/his very constructive and enlightening suggestions, comments and even critics! This work was supported by the World Laboratory for Cosmology And Particle Physics (WLCAPP), http://wlcapp.net/.



\begin{thebibliography}{99}

\bibitem{hf13a} M. J. Giannoni and P. Quentin, Phys. Rev. C {\bf 21}, 2076 (1980).

\bibitem{hf13b} M. Beiner, H. Flocard, N. Van Giai, and P. Quentin, Nucl. Phys. A {\bf 238}, 29 (1975).

\bibitem{hf13c} D. Vautherin and D. Brink, Phys. Rev. C {\bf 5}, 626 (1972).

\bibitem{shortcoming} R.K. Su and H.Q. Song, Phys. Rev. C {\bf 37}, 1770 (1988).

\bibitem{Tawfik:2014eba} Abdel Nasser Tawfik, {\it ''Equilibrium statistical-thermal models in high-energy physics''}, 
Int. J. Mod. Phys. A {\bf 29}, 1430021 (2014).

\bibitem{Tawfik:2011sh} A. Tawfik and  T. Harko, 
Phys. Rev. D {\bf 85}, 084032 (2012). 

\bibitem{Tawfik:2011mw} A. Tawfik, 
Annalen Phys. {\bf 523}, 423 (2011). 

\bibitem{Tawfik:2010pm} A. Tawfik, M. Wahba, H. Mansour and T. Harko, 
Annalen Phys. {\bf 522}, 912  (2010). 

\bibitem{Tawfik:2010ht} A. Tawfik, 
Can. J. Phys. {\bf 88}, 825 (2010). 

\bibitem{Tawfik:2010bm} A. Tawfik, M. Wahba, H. Mansour and T. Harko, 
Annalen Phys. {\bf 523}, 194  (2011). 

\bibitem{Tawfik:2009mk} A. Tawfik, T. Harko, H. Mansour and M. Wahba, 
Uzbek J. Phys. {\bf 12},  316   (2010). 

\bibitem{Tawfik:2010mb} A. Tawfik and M. Wahba, 
 Annalen Phys. {\bf 522}, 849  (2010). 

\bibitem{gupta72} V. Gupta, 
Lett. Nuovo Cim. {\bf 5S2}, 108 (1972).

\bibitem{star1a} Xiaofeng Luo [STAR Collaboration], Acta Phys. Polon. Supp. {\bf 5}, 497 (2012).

\bibitem{star1b} Xiaofeng Luo [STAR Collaboration], J. Phys. Conf. Ser. {\bf 316}, 012003 (2011).

\bibitem{star1c} M.M. Aggarwal {\it et al.} [STAR Collaboration], Phys. Rev. Lett. {\bf 105}, 022302 (2010). 

\bibitem{star2a} T. J. Tarnowsky [STAR Collaboration], J. Phys. G {\bf 38}, 124054 (2011).

\bibitem{star2b} T. K. Nayak [STAR Collaboration]  Nucl. Phys. A {\bf 830}, 555C (2009).

\bibitem{lqcd1a} C. Miao [RBC-Bielefeld Collaboration], 
Nucl. Phys. A {\bf 830}, 705C (2009). 

\bibitem{lqcd1} R.V. Gavai, S. Gupta  and S. Mukherjee, Phys. Rev. D {\bf 71}, 074013 (2005).

\bibitem{lqcd2} M. Cheng,  {\it et al.}, Phys. Rev. D {\bf 77}, 014511 (2008).

\bibitem{endp1} C. Athanasiou, K. Rajagopal and M. Stephanov,  
Phys. Rev. D {\bf 82}, 074008 (2010). 

\bibitem{endp5} M. Cheng {\it et al.} 
Phys. Rev. D {\bf 79}, 074505 (2009). 

\bibitem{qcdlike} B. Stokic {\it et al.}, 
Phys. Lett. B {\bf 673}, 192 (2009). 

\bibitem{latqcd1} R. V. Gavai and S. Gupta, 
Phys. Lett. B {\bf 696}, 459 (2011). 

\bibitem{cw1} X. Luo [STAR Collaboration],  J. Phys. Conf. Ser. {\bf 316}, 012003  (2011). 

\bibitem{cw2}  N. R. Sahoo {\it et al.}, Phys. Rev. C {\bf 87}, 044906 (2013). 


\bibitem{skellm} J. G. Skellam,  
J. Royal Stat. Society A, {\bf 109}, 296 (1946).

\bibitem{Sahoo} N. R. Sahoo, 42nd International Symposium on Multiparticle Dynamics (ISMD 2012)	17-21 Sep. 2012, Kielce-Poland, Acta Phys. Polon. B Proc. Suppl. {\bf 6}, 437 (2013). 

\bibitem{jean2006} J. Cleymans, H. Oeschler, K. Redlich and S. Wheaton, Phys. Rev. C {\bf 73}, 034905 (2006). 

\bibitem{fodor2012} Sz. Borsanyi  {\it et al.}, 
JHEP {\bf 1208}, 053 (2012). 

\bibitem{nxu} P. Garg, D.K. Mishra, P.K. Netrakanti, B. Mohanty, A.K. Mohanty, B.K. Singh, N. Xu, 
Phys. Lett. B {\bf 726}, 691-696  (2013). 

\bibitem{lqcd4} S. Borsanyi, G. Endrodi, Z. Fodor, S.D. Katz, S. Krieg, C. Ratti and K.K. Szabo, JHEP {\bf 1208}, 053 (2012).

\bibitem{book}  M. Le Bellac, F. Mortessagne and G. G. Batrouni, {\it ''Equilibrium and Non-Equilibrium Statistical Thermodynamics''}, (Cambridge University Press, New York, 2004).

\bibitem{Tawfik:2006yq} A. Tawfik, 
Indian J. Phys. {\bf 85}, 755 (2011). 

\bibitem{Tawfik:2006dk} A. Tawfik, 
Indian J. Phys. {\bf 86}, 641 (2012). 

\bibitem{Tawfik:2012zp} A. Tawfik, E. Gamal and A.G. Shalaby, {\it ''Particle Production at RHIC and LHC Energies''},  
1209.5379 [hep-ph].

\bibitem{Tawfik:2012si} A. Tawfik, 
Adv. High Energy Phys. {\bf 2013}, 574871 (2013). 

\bibitem{Tawfik:2008ii} A. Tawfik, 
Indian J. Phys. {\bf 86}, 1139 (2012). 

\bibitem{Tawfik:2012ty} A. Tawfik and H. Magdy, 
Int. J. Mod. Phys. A {\bf 29}, 1450152 (2014).  

\bibitem{Karsch:2003vd} F.~Karsch, K.~Redlich and A.~Tawfik, 
Eur.~Phys.~J.~C {\bf 29},~549~(2003).

\bibitem{Karsch:2003zq} F.~Karsch, K.~Redlich and A.~Tawfik, 
Phys.~Lett.~B {\bf 571},~67~(2003).

\bibitem{Redlich:2004gp} K.~Redlich, F.~Karsch and A.~Tawfik, 
J.~Phys.~G {\bf 30},~S1271~(2004). 

\bibitem{Tawfik:2004sw} A. Tawfik, 
Phys. Rev. D {\bf 71}, 054502 (2005).


\bibitem{Tawfik:2013eua} A. Tawfik, 
 Phys. Rev. C {\bf 88}, 035203 (2013). 

\bibitem{Tawfik:2013dba} A. Tawfik, 
Nucl. Phys. A {\bf 922}, 225 (2014). 

\bibitem{Tawfik:2013pd} A. Tawfik, H. Magdy and E. Gamal, 
Ukr. J. Phys. {\bf 58}, 933 (2013).

\bibitem{Tawfik:2005qn} A. Tawfik, 
Nucl. Phys. A {\bf 764}, 387 (2006). 

\bibitem{formt} R. Chatterjee and D. K. Srivastava, 
Nucl. Phys. A {\bf 830}, 503C (2009).

\end{thebibliography}
\end{document}